\documentclass[11pt]{article}

\usepackage[T1]{fontenc}
\usepackage[utf8]{inputenc}
\usepackage{lmodern}
\usepackage[margin=1in]{geometry}
\usepackage{amsmath,amssymb,bm}
\usepackage{graphicx}
\usepackage{booktabs}
\usepackage{tabularx}
\usepackage{array}
\usepackage{authblk}
\usepackage{natbib}
\usepackage[colorlinks=true,linkcolor=blue,citecolor=blue,urlcolor=blue]{hyperref}

\newcommand{\bs}[1]{\boldsymbol{#1}}
\newcolumntype{C}{>{\centering\arraybackslash}X}

\title{Distribution and Evolution of the Debris Cloud from the Fragmentation of Intelsat 33E}
\author[1]{Peng Shu\thanks{Corresponding author: \href{mailto:shupeng@ynao.ac.cn}{shupeng@ynao.ac.cn}}}
\author[2,4]{Meng Zhao}
\author[1,3]{Yu-Yan Wu}
\author[2,4]{Zhen Yang}
\author[1]{Yu-Qiang Li}
\affil[1]{Yunnan Observatories, Chinese Academy of Sciences, Kunming 650011, China}
\affil[2]{National University of Defense Technology, Changsha 410077, China}
\affil[3]{University of Chinese Academy of Sciences, Beijing 100049, China}
\affil[4]{State Key Laboratory of Space System Operation and Control, Changsha 410077, China}
\date{}

\begin{document}

\maketitle

\begin{abstract}
The breakup of Intelsat 33E on 19 October 2024 posed a potential risk to satellites in the Geostationary Earth Orbit (GEO). This study analyzes the evolution and distribution of these fragments using a probabilistic approach. The initial distribution of the fragments, derived from the NASA Standard Breakup Model, indicates the generation of 4,393 fragments larger than 1 centimeter. The spatial propagation of these fragments is modeled analytically in the Earth-Centered Earth-Fixed reference frame, showing the formation of high-density ring structures in the equatorial plane from 24 hours to 28 days after the breakup. The orbits of 36 cataloged fragments are retrieved and compared with the probability density. Furthermore, Monte Carlo simulations validate the probabilistic model and highlight its efficiency in capturing low-probability events. Collision risks to other GEO satellites are assessed, showing that the top 10\% of satellites encounter a collision probability of up to $10^{-8}$ after 28 days. Satellites near the equatorial plane are at higher risk, whereas those with higher inclinations are less affected. These findings underscore the need for enhanced monitoring and mitigation strategies for GEO breakup events, given the challenges in detecting small fragments.
\end{abstract}


\section{Introduction} \label{sect:intro}

Many valuable satellites are deployed in the Geostationary Earth Orbit (GEO) due to their ability to maintain a fixed position relative to the Earth's surface.
However, the majority of these satellites are confined to a narrow orbital ring above the equator, typically separated by only $\pm 0.1^\circ$ in longitude.
Furthermore, the growing number of GEO satellites has exceeded the availability of designated slots, forcing multiple satellites to share a single longitudinal position \citep{li2014Collocation}.
The high concentration of satellites in the GEO region renders it particularly vulnerable to space debris \citep{shu2025Shortterm}, a situation further compounded by the absence of natural mechanisms for debris removal \citep{yasaka1992Breakup,mei2022Hybrid}.
Recently, Intelsat 33E experienced an anomaly on 19 October 2024 that caused a loss of power and service, and Intelsat reported on 21 October 2024 that the event had resulted in the total loss of the satellite \citep{intelsat2024Loss}. The event was then identified as an on-orbit fragmentation in NASA's Orbital Debris Quarterly News \citep{nasa2025Two}, highlighting the potential for significant pollution of the GEO region.

The primary method for detecting objects in geostationary orbit is optical imaging, as radar effectiveness decreases significantly at such distances \citep{sun2015Algorithms,jiang2022Space}.
However, tracking fragments from breakup events poses a greater challenge, since most are too small to be detected reliably \citep{jiang2022Faint,liu2023Datadriven}.
Typically, only objects larger than 1 meter are cataloged by the US Space Surveillance Network \citep{luo2019FocusGEO}.
The population of GEO objects may exceed the cataloged count by a factor of four, as indicated by a European Space Agency (ESA) survey \cite{hanada2013Effective}.
As of February 2026, only 36 cataloged fragments from Intelsat 33E have been published by the US Space Surveillance System, which is a small fraction of the expected population from such an event.
Public optical tracking reports indicate that the actual fragment cloud is substantially larger; for example, ExoAnalytic Solutions reported more than 700 debris pieces by 10 December 2024, even though only a much smaller subset had cataloged orbital elements available publicly \citep{rainbow2024Intelsat33e}.
Nevertheless, the hazard of untracked fragments is not negligible due to their extremely high speeds \citep{oltrogge2018Comprehensive}.

Dynamic models are a powerful tool to provide valuable insights into the evolution and risk of GEO debris, particularly when observations are limited or incomplete.
Early researchers often utilized simple Monte Carlo methods to simulate the breakup fragments in GEO \citep{yasaka1992Breakup,yasaka1994Remarks}.
However, constrained by the computational resources available at the time, these models incorporated significant simplifications and typically operated with a limited number of samples.

Nowadays, some high-fidelity models have been developed to better simulate the evolution of space debris.
\citet{rossi2009New} developed the Space Debris Mitigation program (SDM) to analyze the debris environment.
\citet{lewis2012Synergy} developed the Debris Analysis and Monitoring Architecture for the Geosynchronous Environment (DAMAGE) to study the mitigation and removal of space debris.
The Harbin Institute of Technology in China established the Space Debris Environment Engineering Models (SDEEM), which simulates debris from Low Earth Orbit (LEO) to GEO \citep{liu2024Space}.
The latest version of SDEEM incorporates interfaces for modeling large constellations and sudden breakup events.
Beyond the models developed in academia, specialized aerospace agencies have also established their own models for engineering purposes.
NASA developed the Orbital Debris Engineering Model (ORDEM) as the primary tool to provide a timely, validated model of the human-made orbital debris environment \citep{matney2019NASA}.
ORDEM generates a set of representative fragments from breakup events, and then propagates them to the future moment.
Similarly, ESA developed the Meteoroid and Space Debris Terrestrial Environment Reference (MASTER) model \citep{krisko2015ORDEM}.
The foundational modeling philosophy of MASTER involves simulating all known debris-generating events in space to establish a synthetic population.
This population can then be used to derive the corresponding space debris flux for a given satellite mission, thereby enabling risk assessment.
These models characterize the evolution of debris fragments by generating random samples and propagating them forward in time.
However, due to stochastic variability, each simulation may yield different outcomes.
Thus, repeated simulations are necessary to draw statistically reliable conclusions \citep{rossi2016Analysis}.

By representing fragments through probability distributions, continuous methods have been introduced to model the evolution of breakup debris.
Letizia et al. proposed an analytical model to describe the evolution of fragments based on the continuity equation \citep{letizia2015Analytical,letizia2018Extension}.
Subsequently, \citet{frey2021Transformation} developed a fully probabilistic framework that transforms initial fragment distributions into a continuous space.
More recently, \citet{wen2024Modeling} derived the medium-term distribution of fragments using inverse mappings of three orbital elements.

This work presents a high-resolution analysis of the post-breakup evolution of the Intelsat 33E debris population, with the goal of quantifying the secondary collision risk to other satellites.
The initial fragments were described using a probability density function (PDF) derived from the NASA SBM.
The PDF of the entire population was then propagated over time using a fully probabilistic method.
The rest of the paper is organized as follows.
Section \ref{sect:breakup} describes the details of the fragments generated from the breakup of Intelsat 33E.
Section \ref{sect:evolution} describes the distribution of fragments over time.
Section \ref{sect:collision} evaluates the risk of secondary collisions to other GEO satellites.
And the conclusions are drawn in Section \ref{sect:conclusion}.

\section{Fragments Generated from the Breakup} \label{sect:breakup}

Intelsat 33E was a high throughput communications satellite manufactured by Boeing on the Boeing 702MP platform \citep{romantisIS33e}.
It orbited the Earth at a height of 35,780 km.
Launched on 24 August 2016 with a satellite mass of 6575 kg, it was positioned at 60$^\circ$E longitude and had an expected operational lifespan of over 15 years \citep{romantisIS33e}.
On 19 October 2024, the satellite experienced an anomaly that caused a loss of power and service; on 21 October 2024, Intelsat reported that the event had resulted in the total loss of the satellite and that a Failure Review Board had been convened to investigate the cause \citep{intelsat2024Loss}. Intelsat later disclosed that most affected services had been restored using other satellites in its fleet and third-party satellites \citep{intelsat2025AnnualReport}.

A total of 36 fragments from Intelsat 33E have been cataloged as of February 2026, according to the Two-Line Element (TLE) data from the Space Track Website \footnote{\url{https://www.space-track.org/}}.
It should be noted, however, that most fragments generated by a breakup event are too small to be detected \citep{nafi2020Practical}. This cataloged count should therefore not be interpreted as the total population. In this paper, we focus on the distribution and evolution of all possible fragments generated from the breakup of Intelsat 33E.

A number of models have been developed to describe the fragmentation of satellites, including the NASA Standard Breakup Model (SBM) \citep{krisko2011Proper,johnson2001NASAs}, the IMPACT model by The Aerospace Corporation \citep{mains2022IMPACT}, the Collision Simulation Tool Solver (CSTS) from the University of Padova \cite{olivieri2023Simulations}, and the Spacecraft Breakup Model developed by the China Aerodynamics Research and Development Center (CARDC-SBM) \citep{yao2024Generation}.
Among these, the NASA SBM is a widely adopted model for predicting the fragments generated from the breakup of satellites.
The characteristics of the fragments, such as size, mass, and velocity, are described by several probability density functions.
In this study, the NASA SBM was employed to determine the initial fragment distribution resulting from the breakup of Intelsat 33E.

\subsection{The Initial PDFs of Fragments} \label{sect:init_pdfs}

The number of fragments from a breakup event is governed by the power law distribution \citep{krisko2011Proper}:
\begin{equation}
    N (L) = S * 6 * (L[m]/1[m])^{-1.6} \label{eq:power}
\end{equation}
where $L$ is the characteristic length of the fragments, $S$ is a unitless factor that depends on the explosive body, and $N$ is the number of fragments with a size larger than $L$.
Thus, the probability density function of the number of fragments is:
\begin{equation}
    p_L(L) = \frac{1}{N}*S * 9.6 * (L[m]/1[m])^{-2.6} \label{eq:pl}
\end{equation}

The distribution of the area-to-mass ratio $A/m$ of fragments is governed by a conditional probability density function relative to the characteristic length $L$:
\begin{equation}
    p_{\chi|\iota}(\chi|\iota) = \sum_{i=1}^{2} \alpha_i(\iota) * \mathcal{N}\left(\mu_{\chi}^{(i)}(\iota), \sigma_{\chi}^{(i)}(\iota)\right) \label{eq:pchilmd}
\end{equation}
where $\chi = \log_{10}(A/m)$ is the logarithm of the area-to-mass ratio, $\iota = \log_{10}(L)$ is the logarithm of the characteristic length, and $\alpha_i$, $\mu_{\chi}^{(i)}$, and $\sigma_{\chi}^{(i)}$ are coefficients determined by the explosive body \citep{johnson2001NASAs}.

At the same time, the distribution of fragment velocity is described by a conditional probability density function relative to the area-to-mass ratio $A/m$:
\begin{equation}
    p_{u|\chi}(u|\chi) = \mathcal{N}\left(\mu_{u}^{\chi}, \sigma_{u}^{\chi}\right) \label{eq:puchi}
\end{equation}
where $u = \log_{10}(\Delta v)$ is the logarithm of the ejection velocity, with a mean of $\mu_{u} = 0.2\chi +1.85$ and a standard deviation of $\sigma_{u} = 0.4$.
By combining Equations \eqref{eq:pl}, \eqref{eq:pchilmd}, and \eqref{eq:puchi}, we obtain the marginal probability density function of the ejection velocity as follows:
\begin{equation}
    p_{u}(u) = \int_{\chi_{\min}}^{\chi_{\max}} p_{u|\chi}(u|\chi) \int_{\iota_{\min}}^{\iota_{\max}} p_{\chi|\iota}(\chi|\iota) p_{\iota}(\iota) \mathrm{d}\iota \mathrm{d}\chi \label{eq:pu}
\end{equation}

\begin{figure}
    \centering
    \includegraphics[width=\textwidth]{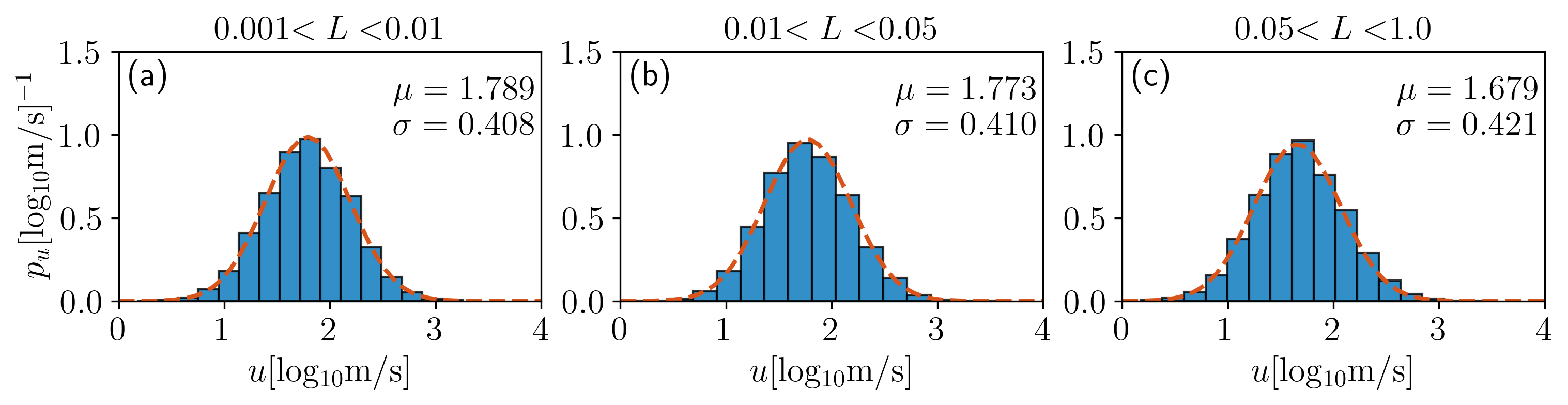}
    \caption{The initial distribution of fragments velocity}
    \label{fig:initvel}
\end{figure}

Figure \ref{fig:initvel} presents the PDF of the ejection velocity of fragments generated from the breakup of Intelsat 33E.
The fragments are then divided into three groups based on their characteristic lengths.
The NASA SBM predicts that 184,533 fragments with a size larger than 0.001 m are generated, of which 4,393 fragments have a size larger than 0.01 m.
The histograms in Figure \ref{fig:initvel} are obtained by random sampling from these conditional PDFs, while the dashed line represents the marginal PDF computed using Eq. \eqref{eq:pu}.
The sampling results suggest that the distribution of $u$ approximately follows a normal distribution, with a mean of approximately 1.7 and a standard deviation of 0.4.
The integrated marginal PDF $p_u$ is consistent with the sampling results, which means that Eq. \eqref{eq:pu} is an accurate description of the distribution of the ejection velocity.

\subsection{The Orbital Elements of Fragments} \label{sect:orbital_elements}

The NASA SBM assumes that the ejection direction of the breakup debris is isotropic.
Then the PDF of the initial velocity vector of fragments is:
\begin{equation}
    p_{\bs{v1}}(\bs{v_1}) = \frac{p_{\Delta v}(\bs{v_1} - \bs{v}_0)}{4\pi \|\bs{v_1} - \bs{v}_0\|^2} \label{eq:pv}
\end{equation}
where $\bs{v}_1$ is the velocity of the fragment right after the breakup time, $\bs{v}_0$ is the velocity of the satellite at the breakup time, and $p_{\Delta v}$ is the probability density function derived from Eq. \eqref{eq:pu}.

\begin{figure}[h]
    \centering
    \includegraphics[width=0.5\textwidth]{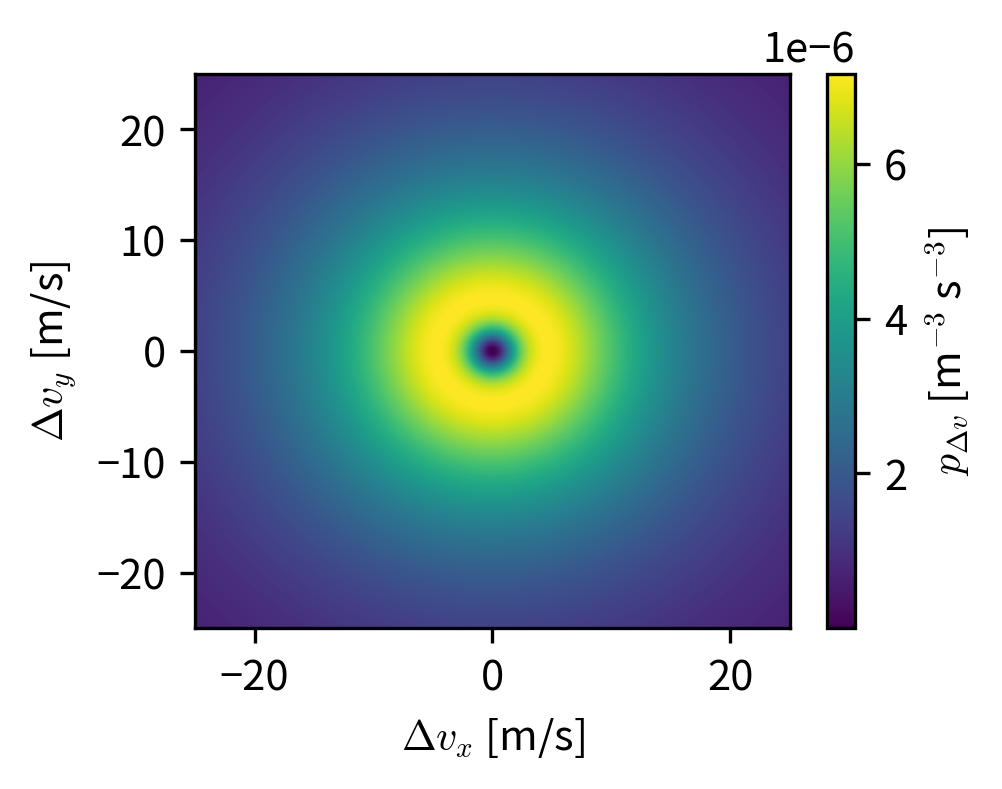}
    \caption{The probability density of the velocity vector}
    \label{fig:pdf_deltav}
\end{figure}

Figure \ref{fig:pdf_deltav} shows the probability density of $\Delta \bs{v}$ within the equatorial plane.
Most fragments are predicted to have a $\Delta \bs{v}$ between -20 m/s and 20 m/s.
Nevertheless, just a few fragments are expected to have a small velocity increment less than 1 m/s.
This implies that most fragments will gradually drift away from the satellite's original orbital position.

In addition, the initial position of each fragment is assumed to coincide with that of the satellite at the time of the breakup.
Based on this assumption, the initial orbital elements of the fragments can be determined from their initial state vectors.

\begin{figure}
    \centering
    \includegraphics[width=0.5\textwidth]{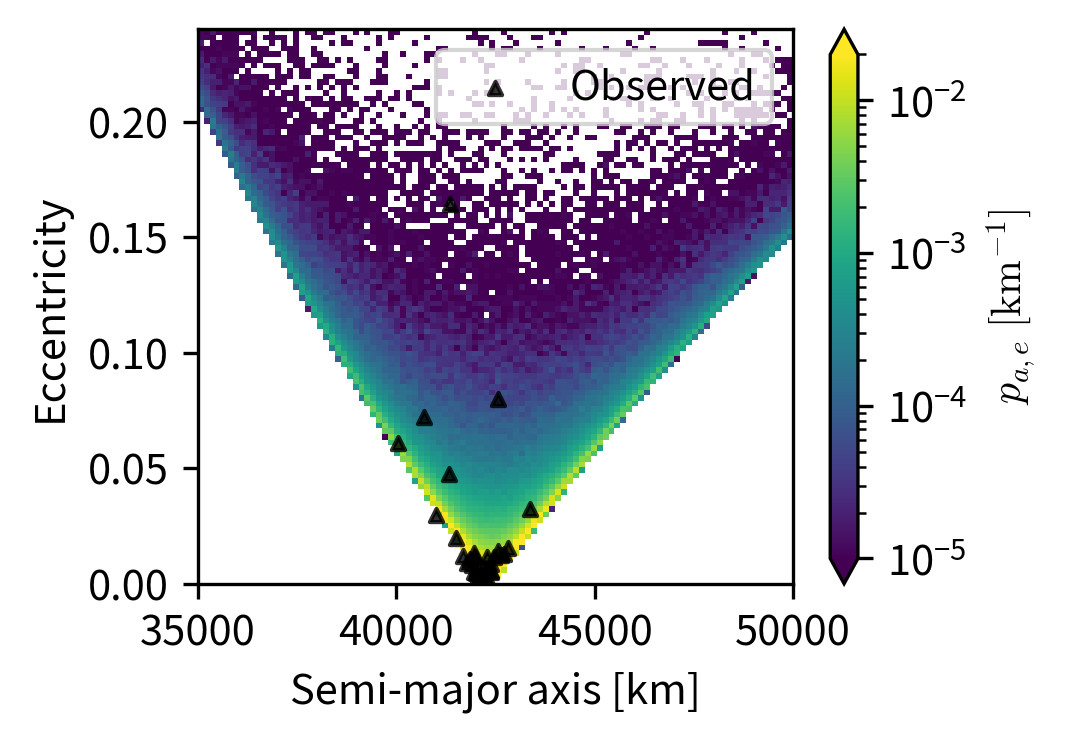}
    \caption{The semi-major axis and eccentricity of fragments}
    \label{fig:ae}
\end{figure}

\begin{figure}
    \centering
    \includegraphics[width=0.5\textwidth]{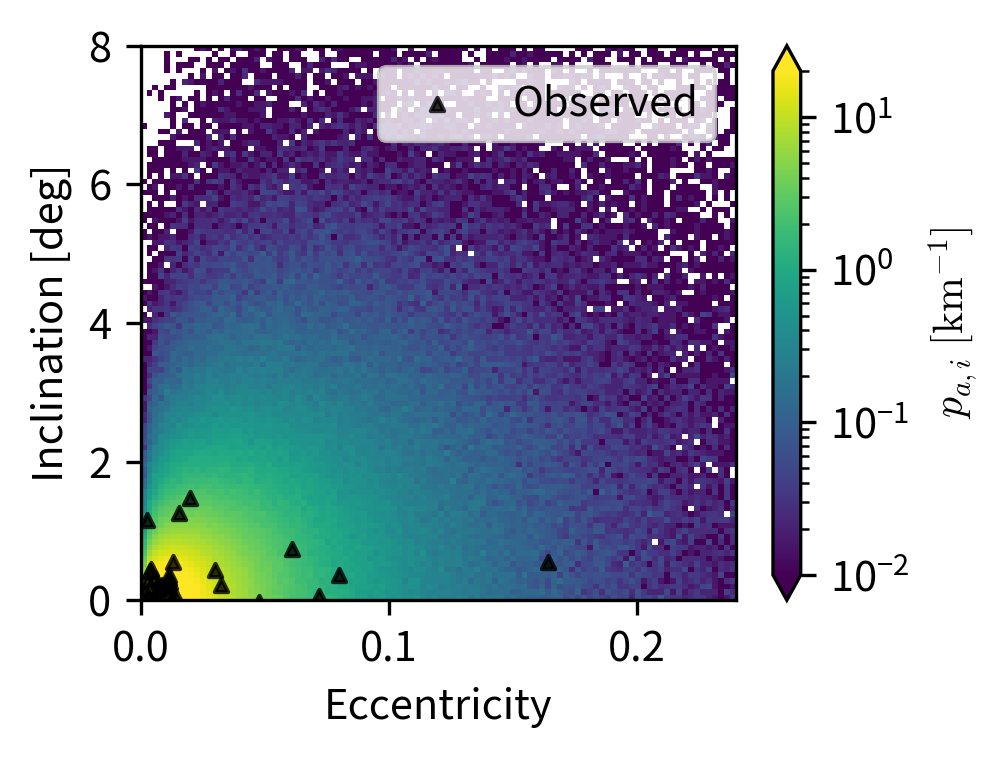}
    \caption{The orbital inclination and eccentricity of fragments}
    \label{fig:ei}
\end{figure}

The colormap in Fig. \ref{fig:ae} illustrates the distribution of semi-major axis and eccentricity of fragments resulting from the breakup of Intelsat 33E.
According to the model, the majority of fragments have a semi-major axis between 40,000 km and 60,000 km and an eccentricity of less than 0.5.
Meanwhile, the colormap in Fig. \ref{fig:ei} depicts the inclination distribution of the fragments, which is predicted to be predominantly below $2^\circ$.

The black dots in Figures \ref{fig:ae} and \ref{fig:ei} represent the 36 observed fragments retrieved from TLE data.
These observed fragments are generally consistent with the predicted distribution, except for a few outliers exhibiting higher eccentricity.

\section{Evolution of Fragments Distribution} \label{sect:evolution}

\subsection{The Probability Density of Fragments State Vector} \label{sect:pdf_state}

At the moment of breakup, all fragments are assumed to have the same position as Intelsat 33E, with the probability density function (PDF) of their state vector given by:
\begin{equation}
    p_{\bs{x1}}(\bs{x_1}) = \delta(\bs{r_1} - \bs{r}_0)p_{\bs{v1}}(\bs{v_1}) \label{eq:px}
\end{equation}
where $\bs{x_1} = (\bs{r_1}, \bs{v_1})$ is the initial state vector of fragments, $\delta(\cdot)$ is the Dirac delta function, $\bs{r}_0$ is the position of Intelsat 33E at the breakup time, and $p_{\bs{v1}}$ is given by Eq. \eqref{eq:pv}.

For an arbitrary fragment, the position and velocity can be expressed as a function of time:
\begin{equation}
\begin{aligned}
    \bs{r}_2 &= \varphi_r(t, \bs{r_1}, \bs{v_1}) \\
    \bs{v}_2 &= \varphi_v(t, \bs{r_1}, \bs{v_1}) \label{eq:statetrans}
\end{aligned}
\end{equation}
where $t$ is the time elapsed since the breakup.

The probability density of the state vector of fragments at time $t$ can then be derived using change of variables formula \citep{shu2023Impact}:
\begin{equation}
    p_{\bs{x}2}(\bs{x}_2) = \delta\left(\varphi_r^{-1}(t, \bs{r}_2, \bs{v}_2) - \bs{r}_0\right) p_{\bs{v1}}\left(\varphi_v^{-1}(t, \bs{r}_2, \bs{v}_2)\right) \left| \det\left(\frac{\partial \bs{x}_2}{\partial \bs{x_1}}\right)\right|^{-1} \label{eq:px2}
\end{equation}
The equation
\begin{equation}
    \varphi_r^{-1}(t, \bs{r}_2, \bs{v}_2) - \bs{r}_0 = 0 \label{eq:bvp}
\end{equation}
may have multiple solutions, i.e., there are multiple orbits linking the position $\bs{r}_0$ at the breakup time and the position $\bs{r}_2$ at time $t$.
We denote these solutions as $\bs{v}_{2*}^{(i)}$, where $i = 1, 2, \ldots, m$ and $m$ is the number of solutions.
Then the delta function can be expanded as:
\begin{equation}
    \delta(\varphi_r^{-1}(t, \bs{r}_2, \bs{v}_2) - \bs{r}_0) = \sum_{i=1}^{m} \delta\left(\bs{v}_2 - \bs{v}_{2*}^{(i)}\right) \left| \det\left(\frac{\partial \varphi_r^{-1}}{\partial \bs{v}_2}\right)\right|^{-1} \label{eq:delta_expand}
\end{equation}

Thus, the probability density function of the state vector of fragments at time $t$ can be rewritten as:
\begin{align}
    p_{\bs{x}2}(\bs{x}_2) &= \sum_{i=1}^{m} \delta\left(\bs{v}_2 - \bs{v}_{2*}^{(i)}\right) p_{\bs{v1}}\left(\varphi_v^{-1}(t, \bs{r}_2, \bs{v}_{2*}^{(i)})\right) \left| \det\left(\frac{\partial \varphi_r^{-1}}{\partial \bs{v}_2}\right)\right|^{-1} \left| \det\left(\frac{\partial \bs{x}_2}{\partial \bs{x_1}}\right)\right|^{-1} \nonumber \\
    &= \sum_{i=1}^{m} \delta\left(\bs{v}_2 - \bs{v}_{2*}^{(i)}\right) p_{\bs{v1}}\left(\varphi_v^{-1}(t, \bs{r}_2, \bs{v}_{2*}^{(i)})\right) \left| \det\left(\frac{\partial \bs{r}_2}{\partial \bs{v}_1}\right)\right|^{-1} \label{eq:px2_final}
\end{align}
Equation \eqref{eq:px2_final} indicates that the PDF of the state vector of fragments at time $t$ is a sum of several terms, where each term corresponds to a solution of the boundary value problem defined by $\bs{r_0}$, $\bs{r}_2$, and $t$.

Furthermore, the probability density function of fragments position at time $t$ can be obtained by integrating Eq. \eqref{eq:px2_final} over the velocity space:
\begin{align}
    p_{\bs{r}2}(\bs{r}_2) &= \int_{\mathbb{R}^3} p_{\bs{x}2}(\bs{x}_2) \mathrm{d}\bs{v}_2 \nonumber \\
    &= \sum_{i=1}^{m} p_{\bs{v1}}\left(\varphi_v^{-1}(t, \bs{r}_2, \bs{v}_{2*}^{(i)})\right) \left| \det\left(\frac{\partial \bs{r}_2}{\partial \bs{v}_1}\right)\right|^{-1} \label{eq:pr2}
\end{align}
Equation \eqref{eq:pr2} represents the probability density distribution of a random fragment, namely, the probability that a single fragment following the initial distribution appears at position $\bs{r}_2$ at time $t$.
If the initial fragment cloud contains $N$ fragments, the corresponding number density can be written as
\begin{equation}
    n_{\bs{r}2}(\bs{r}_2) = N p_{\bs{r}2}(\bs{r}_2)
    \label{eq:nr2}
\end{equation}

\subsection{Fragments Distribution in Earth-Centered Earth-Fixed Frame} \label{sect:frag_ecef}

Given that GEO satellites are stationary relative to the Earth's surface, it is more convenient to analyze the distribution of GEO fragments in the Earth-Centered Earth-Fixed (ECEF) frame.
The ECEF position of a fragment can be computed from its longitude $\lambda$, latitude $\phi$, and height $h$:
\begin{equation}
    \bs{r}_{\mathrm{ECEF}} = (R_E + h) \begin{bmatrix}
        \cos\phi \cos\lambda \\
        \cos\phi \sin\lambda \\
        \sin\phi
    \end{bmatrix} \label{eq:ecef}
\end{equation}
where $R_E$ is the radius of the Earth.
The ECEF position of the fragment can be converted to the ECI position using the rotation matrix:
\begin{equation}
    \bs{r}_{\mathrm{ECI}} = \bs{R}_{\mathrm{E}} \bs{r}_{\mathrm{ECEF}} \label{eq:R_ecef}
\end{equation}
where $\bs{R}_{\mathrm{E}}$ is the rotation matrix from the ECEF frame to the ECI frame at time $t$.
Now, the fragments distribution in the ECEF frame can be obtained by substituting Eq. \eqref{eq:ecef} and \eqref{eq:R_ecef} into Eq. \eqref{eq:pr2}.

\begin{figure}[bp]
    \centering
    \includegraphics[width=0.8\textwidth]{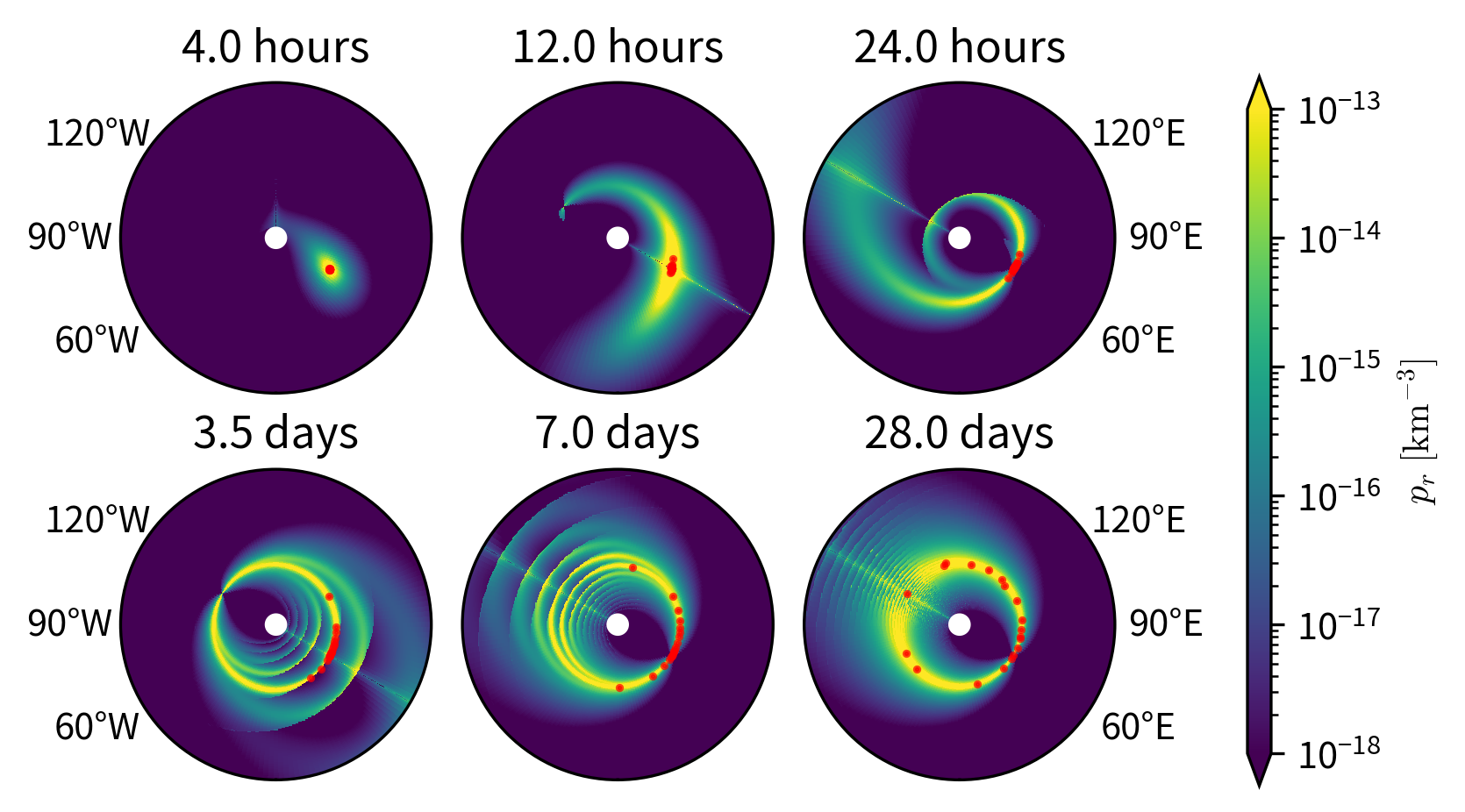}
    \caption{The evolution of fragments in the equatorial plane}
    \label{fig:evolution-geo-polarxy}
\end{figure}

\begin{figure}
    \centering
    \includegraphics[width=0.8\textwidth]{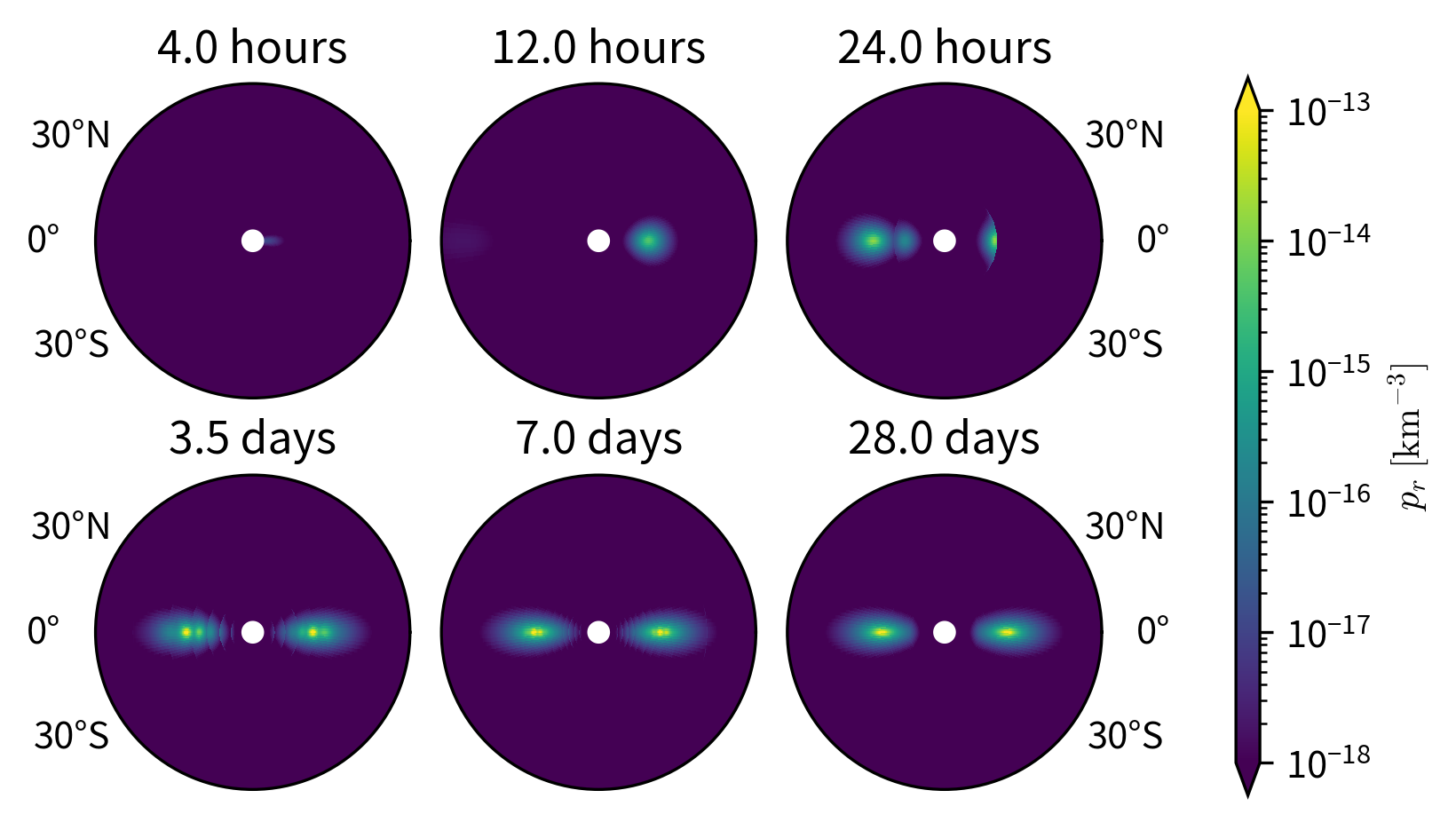}
    \caption{The evolution of fragments in the meridian plane at longitude 150$^\circ$E}
    \label{fig:evolution-geo-polaryz}
\end{figure}

Figures \ref{fig:evolution-geo-polarxy} and \ref{fig:evolution-geo-polaryz} illustrate the evolution of the fragment distribution in the ECEF frame.
The first figure corresponds to the equatorial plane, while the second shows the distribution in the meridian plane at 150$^\circ$E longitude.
The white circular region at the center represents the Earth, with longitude and latitude markings indicated along the outer circumference.
These two figures were generated from the probability density obtained using Eq. \eqref{eq:pr2}.
The colorbar indicates the base-10 logarithm of the probability density of fragments position, calculated using Eq. \eqref{eq:pr2}.
Additionally, the positions of the 36 observed fragments, retrieved from TLE data, are indicated by red dots in Fig. \ref{fig:evolution-geo-polarxy}.

At the time of breakup, all fragments are concentrated at 60$^\circ$E longitude, 0$^\circ$ latitude, and an altitude of 35,780 km.
After 4 hours, the majority of fragments remain near the original breakup location, although a small number are predicted to have drifted to 150$^\circ$E.
By 12 hours, while most fragments stay within a few degrees of the initial position, several have undergone significant longitudinal drift.
The leading fragments have advanced to 120$^\circ$W longitude, whereas the trailing fragments have just passed 0$^\circ$ longitude.
This dispersion arises because fragments experiencing a negative $\Delta v$ acquire a smaller semi-major axis and thus a shorter orbital period, while those with a positive $\Delta v$ attain a larger semi-major axis and a longer orbital period.
As shown in Fig. \ref{fig:evolution-geo-polaryz}, the distribution in the meridian plane also evolves over time, with numerous fragments being ejected away from the equatorial plane due to velocity components perpendicular to the orbital plane.

At 24 hours, the leading fragments have returned to 60$^\circ$E longitude, while the trailing fragments have just passed 120$^\circ$W.
This resulted in the formation of an outward spiral structure in the equatorial plane.
By 3.5 days, the fragments have formed a complete ring around the Earth in the equatorial plane, with the distribution in the meridian plane also becoming more symmetric.
Subsequently, additional rings emerge, and the fragment distribution grows increasingly uniform over time.
Between 12 hours and 28 days, a distinct linear structure is visible in the equatorial plane.
This line consistently aligns toward 120$^\circ$W every 24 hours, indicating a higher density of fragments in the direction opposite to the original breakup location.

The red dots in Fig. \ref{fig:evolution-geo-polarxy} represent the distribution of the observed fragments.
Although these observed fragments are consistent with the predicted distribution, many high-density regions show no detected fragments.
This suggests that many fragments from the Intelsat 33E breakup are difficult to detect with ground-based telescopes.

\subsection{Monte Carlo Simulation of Fragments Distribution} \label{sect:mc_simulation}

To validate the results obtained using the probabilistic method, we also performed a Monte Carlo simulation of the fragments distribution.
A total of 10 million fragments were generated based on the initial PDFs described in Section \ref{sect:init_pdfs}.
They were then propagated to a specified time using a high-precision numerical integrator.

\begin{figure}
    \centering
    \includegraphics[width=0.8\textwidth]{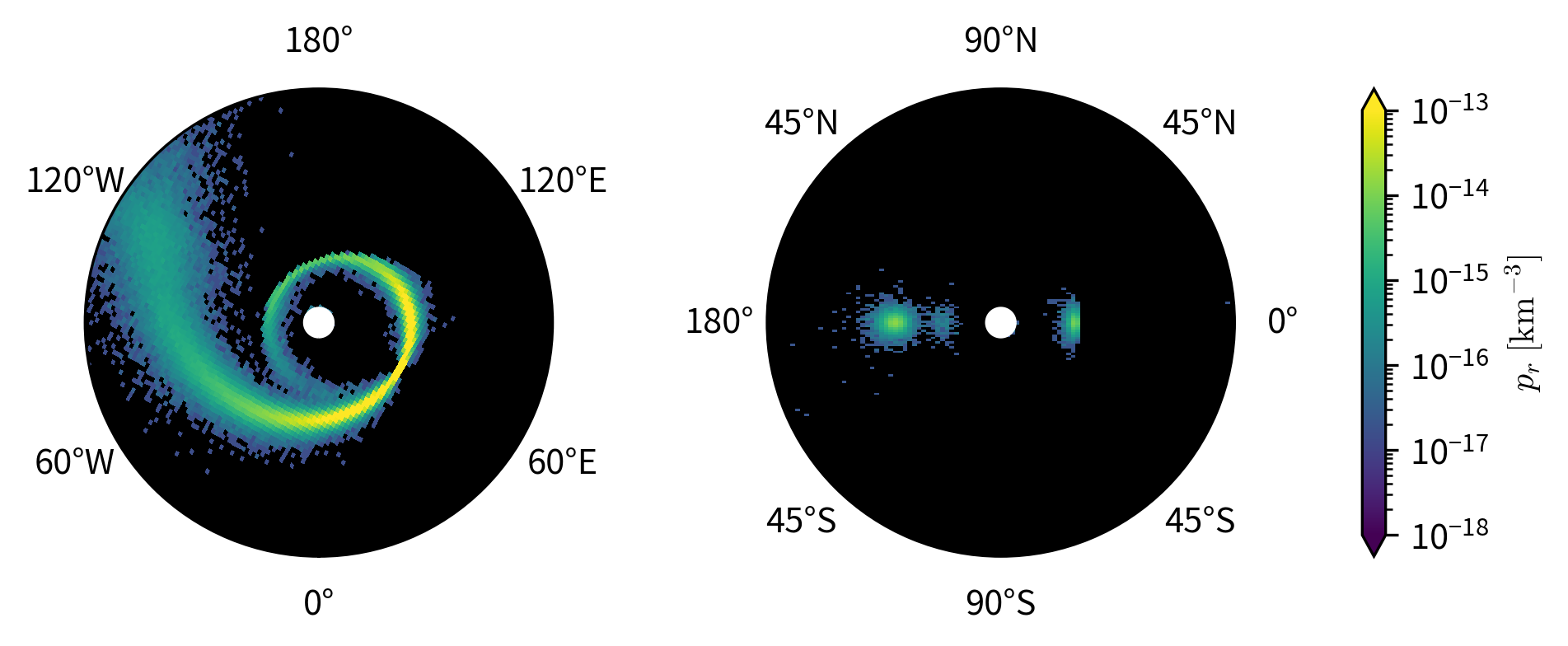}
    \caption{The distribution at 24 hours computed using the Monte Carlo method}
    \label{fig:mc_prop_polar}
\end{figure}

Figure \ref{fig:mc_prop_polar} shows the fragment distribution after 24 hours, computed using the Monte Carlo method.
The probability density was obtained by normalizing the fragment count in each grid cell by both the total number of fragments and the cell area.
The resulting distribution closely matches the analytical results presented in Figures \ref{fig:evolution-geo-polarxy} and \ref{fig:evolution-geo-polaryz}, thus validating the accuracy of the probabilistic method.
Some narrow linear features are less distinct in the Monte Carlo plot because the density is averaged over finite grid cells.
Nevertheless, checks at representative locations are consistent with the analytical results.

On the other hand, certain regions appear black in Fig. \ref{fig:mc_prop_polar}, indicating an absence of samples in these areas.
This occurs because the Monte Carlo method lacks sensitivity to low-probability events \citep{frey2021Transformation}, as capturing a smooth distribution requires a very large number of samples, which is computationally expensive.
In contrast, the density transfer method introduced in Section \ref{sect:pdf_state} can effectively resolve low-probability regions, offering a more efficient alternative for predicting the evolution of fragment clouds.

\section{Collision Probability to GEO Satellites} \label{sect:collision}

\begin{figure}[hbp]
    \centering
    \includegraphics[width=0.4\textwidth]{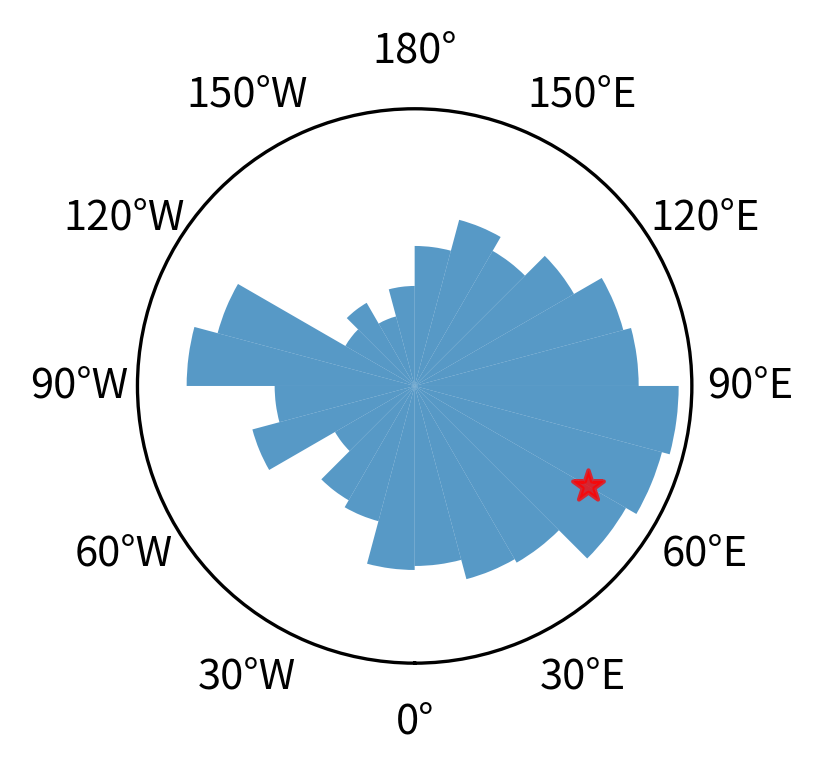}
    \caption{The longitude distribution of GEO satellites}
    \label{fig:geo-longitude}
\end{figure}

The geostationary orbit hosts a population exceeding 1,000 satellites, distributed in a ring with a radius of approximately 42,164 km.
These satellites are concentrated within a few degrees of the equatorial plane, with their longitudes clustered in several specific regions.
Figure \ref{fig:geo-longitude} illustrates this distribution, indicating the most densely populated regions are around 80$^\circ$E and 100$^\circ$W.
The breakup position is marked by a red star.
All assets in this orbital regime face a collision risk with fragments originating from the Intelsat 33E breakup.

The debris flux incident on a satellite surface can be derived from the probability density function of fragments \citep{shu2023Impact}:
\begin{equation}
    F_{\mathrm{in}} = N \int_{\mathbb{R}^3} p_{\bs{x}2}(\bs{r}_s,\bs{v}_2) R\left( -(\bs{v}_2 - \bs{v}_s) \cdot \hat{\bs{n}} \right) \mathrm{d} \bs{v}_2 \label{eq:flux}
\end{equation}
where $N$ is the total number of fragments, $\bs{r}_s$ and $\bs{v}_s$ are the position and velocity of the satellite, respectively, $\hat{\bs{n}}$ is the unit normal vector of the satellite surface, and $R(x)$ is the ramp function defined as:
\begin{equation}
    R(x) = \begin{cases}
        x & x \geq 0 \\
        0 & x < 0
    \end{cases} \label{eq:ramp}
\end{equation}
The number of impacts to a satellite per unit time can then be computed as:
\begin{equation}
    \dot{\eta} (t) = A_c F_{\mathrm{in}} \label{eq:impact_rate}
\end{equation}
where $A_c$ is the cross-sectional area of the satellite.
Finally, the collision probability to the satellite over time $t$ can be computed as:
\begin{equation}
    P(t) =1 - \mathrm{e} ^{-\int_0^t \dot{\eta} (t) \mathrm{d} t} \label{eq:collision_prob}
\end{equation}

\begin{figure}
    \centering
    \includegraphics[width=\textwidth]{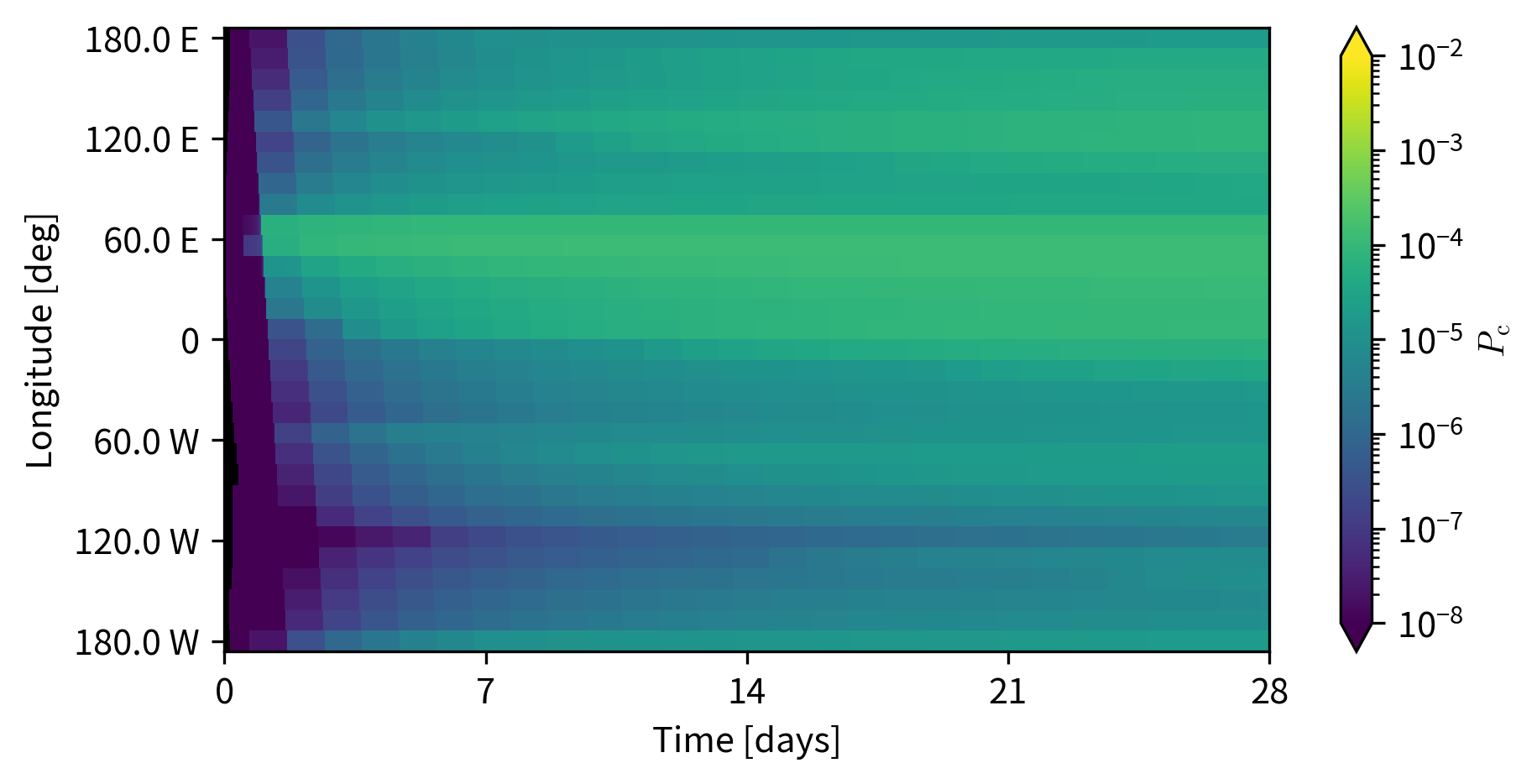}
    \caption{The collision probability to GEO objects}
    \label{fig:pc_lon}
\end{figure}

Figure \ref{fig:pc_lon} illustrates the variation in collision probability with virtual objects over time at different longitudes.
The GEO objects data used in this analysis were downloaded from the Space-Track website\footnote{\url{https://www.space-track.org/}}.
The virtual objects are assumed to follow circular orbits with a radius of 42,164 km and an inclination of 0$^\circ$.
The collision probability is calculated using Eqs. \eqref{eq:flux} to \eqref{eq:collision_prob}, in which the satellite is modeled as a sphere with a radius of 10 m.
The number of fragments, $N$, is taken as 4,393, corresponding to the number of fragments larger than 0.01 m predicted by the NASA SBM.
Thus, only fragments exceeding 0.01 m in size are included in the collision probability assessment.
This collision-risk calculation is therefore associated with a specific assumed fragment population.
If a different size threshold is of interest, the same framework can be applied by replacing $N$ in Eq. \eqref{eq:flux} with the corresponding fragment number.
For example, for fragments larger than 1 mm, one may take $N=184{,}533$ and then evaluate the collision probability using Eq. \eqref{eq:collision_prob}.

During the first 24 hours, the collision probability remains very low across all longitudes, with a maximum value of only $10^{-7}$.
This is attributed to the fact that most fragments remain in close proximity to their original position at this early stage.
Between 1 and 7 days, objects near 60$^\circ$E experience a higher collision risk, as the fragment cloud remains concentrated around the breakup location.
In contrast, the region near 120$^\circ$W exhibits lower risk due to its opposition to the breakup point. From 7 to 28 days, the collision probability increases significantly across all longitudes, with the maximum value reaching $10^{-3}$ and the minimum remaining around $10^{-7}$.

\begin{figure}
    \centering
    \includegraphics[width=0.5\textwidth]{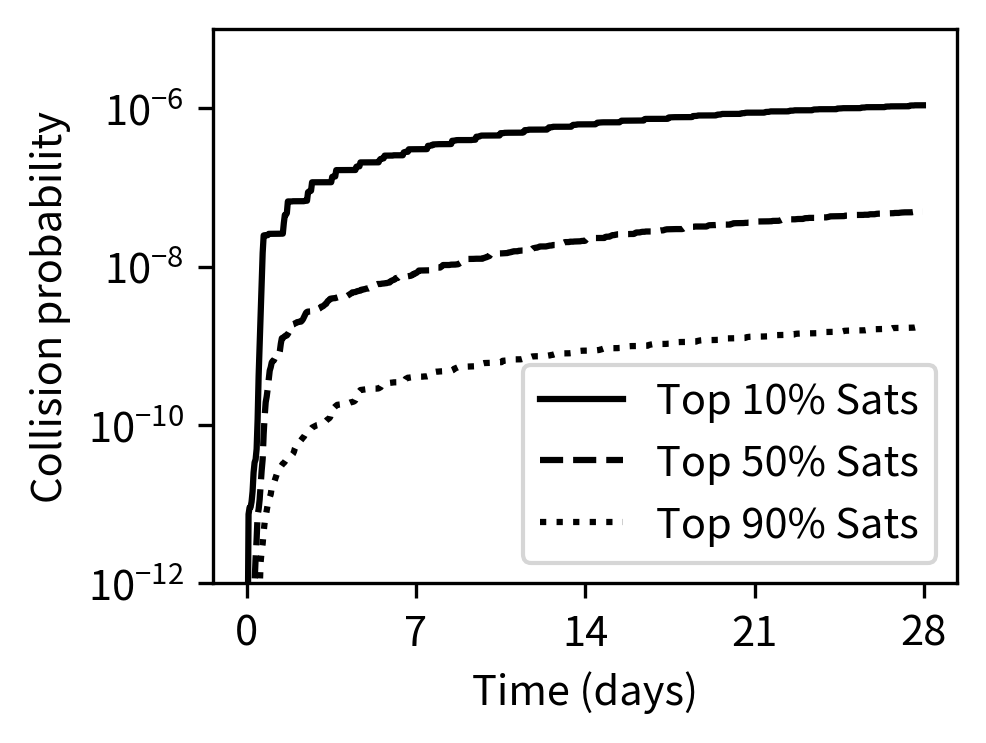}
    \caption{The risk of collision to current GEO satellites}
    \label{fig:geo_sat_risk}
\end{figure}

Figure \ref{fig:geo_sat_risk} illustrates the evolution of collision probability for current GEO satellites over time, with satellite positions obtained from historical TLE data.
The three curves in the figure correspond to the top 10\%, top 50\%, and top 90 percentiles of collision probability, respectively.
The collision probability for the top 10\% of satellites rose rapidly within the first day, reaching a value of $10^{-8}$, and continued to increase gradually to $10^{-6}$ by day 28.
In contrast, the top 90\% of satellites experienced a significantly lower collision probability, which reached approximately $10^{-10}$ in the first week and slowly rose to $10^{-9}$ by the end of the 28-day period.
These values represent impact probability rather than catastrophic damage probability. Although relative velocities in GEO are generally lower than in LEO, small fragments can still threaten vulnerable components such as solar arrays, antennas, and optical surfaces, including those of observation spacecraft \citep{garoli2020Mirrors}; a detailed damage assessment is beyond the scope of this study.

Table \ref{tab:top5} presents the top 5 satellites with the highest collision probabilities at 28 days after the breakup event.
The satellite exhibiting the highest risk is GAOFEN 13, with a collision probability of $1.2 \times 10^{-5}$.
The second-highest risk corresponds to AEHF 4 (USA 288), which has a collision probability of $3.0 \times 10^{-6}$.
This satellite is in a near-geostationary orbit with a slight inclination of 1.6216$^\circ$ and is located at 67.21$^\circ$W longitude.
The remaining three satellites in the top five list are all communication satellites, each with a collision probability exceeding $2 \times 10^{-6}$.

\begin{table}[bp]
    \centering
    \caption{The top 5 satellites with the highest collision probability at 28 days}
    \begin{tabularx}{\textwidth}{CCCCC}
        \toprule
        NORAD ID & Name & $P_c$ & Longitude ($^\circ$) & Inclination ($^\circ$) \\
        \midrule
        46610 & GAOFEN 13 & $1.2 \times 10^{-5}$ & 117.67 $^\circ$E & 0.0415 \\
        43651 & AEHF 4 (USA 288) & $3.0 \times 10^{-6}$ & 67.21 $^\circ$W & 1.6216 \\
        39157 & CHINASAT 11 & $2.9 \times 10^{-6}$ & 97.86 $^\circ$E & 0.0163 \\
        41747 & INTELSAT 36 & $2.04 \times 10^{-6}$ & 68.48 $^\circ$E & 0.0168 \\
        38740 & INTELSAT 20 & $2.02 \times 10^{-6}$ & 68.49 $^\circ$E & 0.0129 \\
        \bottomrule
    \end{tabularx}
    \label{tab:top5}
\end{table}

Table \ref{tab:low5} presents the five satellites with the lowest collision probabilities at 28 days after the breakup event.
The satellite exhibiting the lowest risk is BEIDOU 3 IGSO-3, with a collision probability of only $9.09 \times 10^{-11}$.
All five satellites in this group have relatively high orbital inclinations, ranging from 30$^\circ$ to 55$^\circ$.
This pattern suggests that satellites in highly inclined orbits are subject to a significantly lower collision risk from the fragment cloud generated by the Intelsat 33E breakup.

\begin{table}
    \centering
    \caption{The last 5 satellites with the lowest collision probability at 28 days}
    \begin{tabularx}{\textwidth}{CCCCC}
        \toprule
        NORAD ID & Name & $P_c$ & Longitude ($^\circ$) & Inclination ($^\circ$) \\
        \midrule
        40938 & BEIDOU 20 & $2.35 \times 10^{-10}$ & 105.80 $^\circ$E & 50.16 \\
        634 & SYNCOM 2 & $1.72 \times 10^{-10}$ & 65.83 $^\circ$E & 31.28 \\
        37256 & BEIDOU 7 & $1.31 \times 10^{-10}$ & 114.60 $^\circ$E & 48.65 \\
        40549 & BEIDOU 17 & $9.53 \times 10^{-11}$ & 87.70 $^\circ$E & 50.34 \\
        44709 & BEIDOU 3 IGSO-3 & $9.09 \times 10^{-11}$ & 131.30 $^\circ$E & 55.84 \\
        \bottomrule
    \end{tabularx}
    \label{tab:low5}
\end{table}

\section{Conclusion} \label{sect:conclusion}

The fragment cloud generated by the breakup of Intelsat 33E poses a collision risk to current GEO satellites.
The initial probability density function of fragment velocity is derived from the distributions of characteristic length and area-to-mass ratio of the debris.
The breakup is predicted to produce 4,393 fragments larger than 0.01 m and 184,533 fragments larger than 0.001 m. Most fragments exhibit an eccentricity below 0.5 and an inclination under 2$^\circ$.
The spatial distribution of fragments in the Earth-Centered Earth-Fixed frame is propagated from the initial probability density function.
The results reveal the emergence of high-density rings in the equatorial plane starting 24 hours after the breakup.
Large-sample Monte Carlo simulations confirm that the density transfer method effectively captures low-probability features of the fragment distribution, offering a computationally efficient approach for predicting the evolution of debris clouds.
Collision probabilities with current GEO satellites are evaluated based on the evolved fragment distribution.
After 28 days, the top 10\% of satellites exhibit a collision probability of $10^{-8}$, substantially higher than the $10^{-10}$ probability for the top 90\% of satellites.
This suggests that some GEO satellites are at risk of collision with the fragments.
Even when such impacts are not catastrophic, small debris can still degrade mission performance by damaging exposed and sensitive components, including solar arrays, antennas, and optical surfaces, which is also relevant to observation spacecraft.
Special attention should be paid to GEO breakup events, as the majority of the fragments are difficult to detect.

\bibliographystyle{plainnat}
\bibliography{bibtex}

@string{jan = "Jan"}

@string{feb = "Feb"}

@string{mar = "Mar"}

@string{apr = "Apr"}

@string{may = "May"}

@string{jun = "Jun"}

@string{jul = "Jul"}

@string{aug = "Aug"}

@string{sep = "Sep"}

@string{oct = "Oct"}

@string{dec = "Dec"}

@article{frey2021Transformation,
  title = {Transformation of {{Satellite Breakup Distribution}} for {{Probabilistic Orbital Collision Hazard Analysis}}},
  author = {Frey, Stefan and Colombo, Camilla},
  year = {2021},
  journal = {Journal of Guidance, Control, and Dynamics},
  volume = {44},
  number = {1},
  pages = {88--105},
  issn = {1533-3884},
  doi = {10.2514/1.G004939},
  urldate = {2020-12-17},
  langid = {english}
}

@article{garoli2020Mirrors,
  title = {Mirrors for Space Telescopes: Degradation Issues},
  author = {Garoli, Denis and Rodriguez De Marcos, Luis V. and Larruquert, Juan I. and Corso, Alain J. and Proietti Zaccaria, Remo and Pelizzo, Maria G.},
  year = {2020},
  month = oct,
  journal = {Applied Sciences},
  volume = {10},
  number = {21},
  pages = {7538},
  doi = {10.3390/app10217538},
  urldate = {2026-03-18}
}

@article{hanada2013Effective,
  title = {Effective {{Search Strategy Applicable}} for {{Breakup Fragments}} in the {{Geostationary Region}}},
  author = {Hanada, Toshiya and Uetsuhara, Masahiko and Yanagisawa, Toshifumi and Kitazawa, Yukihito},
  year = {2013},
  journal = {Journal of Spacecraft and Rockets},
  volume = {50},
  number = {4},
  pages = {802--806},
  publisher = {{American Institute of Aeronautics and Astronautics}},
  issn = {0022-4650},
  doi = {10.2514/1.A32228},
  urldate = {2025-09-21}
}

@article{jiang2022Faint,
  title = {Faint {{Space Debris Detection Algorithm Based}} on {{Small Aperture Telescope Detection System}}},
  author = {Jiang, Ping and Liu, Chengzhi and Kang, Zhe and Yang, Wenbo and Li, Zhenwei},
  year = {2022},
  month = sep,
  journal = {Research in Astronomy and Astrophysics},
  volume = {22},
  number = {10},
  pages = {105003},
  publisher = {{National Astromonical Observatories, CAS and IOP Publishing}},
  issn = {1674-4527},
  doi = {10.1088/1674-4527/ac8b5a},
  urldate = {2025-09-22},
  langid = {english}
}

@article{jiang2022Space,
  title = {Space {{Debris Automation Detection}} and {{Extraction Based}} on a {{Wide-field Surveillance System}}},
  author = {Jiang, Ping and Liu, Chengzhi and Yang, Wenbo and Kang, Zhe and Fan, Cunbo and Li, Zhenwei},
  year = {2022},
  month = feb,
  journal = {The Astrophysical Journal Supplement Series},
  volume = {259},
  number = {1},
  pages = {4},
  publisher = {The American Astronomical Society},
  issn = {0067-0049},
  doi = {10.3847/1538-4365/ac458d},
  urldate = {2025-09-22},
  langid = {english}
}

@article{johnson2001NASAs,
  title = {{{NASA}}'s New Breakup Model of Evolve 4.0},
  author = {Johnson, N. L. and Krisko, P. H. and Liou, J. -C. and {Anz-Meador}, P. D.},
  year = {2001},
  month = jan,
  journal = {Advances in Space Research},
  volume = {28},
  number = {9},
  pages = {1377--1384},
  issn = {0273-1177},
  doi = {10.1016/S0273-1177(01)00423-9},
  urldate = {2020-06-03},
  langid = {english}
}

@article{krisko2011Proper,
  title = {Proper Implementation of the 1998 {{NASA}} Breakup Model},
  author = {Krisko, P. H.},
  year = {2011},
  journal = {Orbital Debris Quarterly News},
  volume = {15},
  number = {4},
  pages = {4--5}
}

@article{krisko2015ORDEM,
  title = {{{ORDEM}} 3.0 and {{MASTER-2009}} Modeled Debris Population Comparison},
  author = {Krisko, P. H. and Flegel, S. and Matney, M. J. and Jarkey, D. R. and Braun, V.},
  year = {2015},
  month = aug,
  journal = {Acta Astronautica},
  volume = {113},
  pages = {204--211},
  issn = {0094-5765},
  doi = {10.1016/j.actaastro.2015.03.024},
  urldate = {2020-08-12},
  langid = {english}
}

@article{letizia2015Analytical,
  title = {Analytical {{Model}} for the {{Propagation}} of {{Small-Debris-Object Clouds After Fragmentations}}},
  author = {Letizia, Francesca and Colombo, Camilla and Lewis, Hugh G.},
  year = {2015},
  journal = {Journal of Guidance, Control, and Dynamics},
  volume = {38},
  number = {8},
  pages = {1478--1491},
  doi = {10.2514/1.G000695},
  urldate = {2020-09-01}
}

@article{letizia2018Extension,
  title = {Extension of the {{Density Approach}} for {{Debris Cloud Propagation}}},
  author = {Letizia, Francesca},
  year = {2018},
  month = sep,
  journal = {Journal of Guidance, Control, and Dynamics},
  volume = {41},
  number = {12},
  pages = {2651--2657},
  doi = {10.2514/1.G003675},
  urldate = {2020-09-01}
}

@article{lewis2012Synergy,
  title = {Synergy of Debris Mitigation and Removal},
  author = {Lewis, Hugh G. and White, Adam E. and Crowther, Richard and Stokes, Hedley},
  year = {2012},
  month = dec,
  journal = {Acta Astronautica},
  volume = {81},
  number = {1},
  pages = {62--68},
  issn = {0094-5765},
  doi = {10.1016/j.actaastro.2012.06.012},
  urldate = {2025-09-21}
}

@incollection{li2014Collocation,
  title = {Collocation {{Prototypes}} and {{Strategies}}},
  booktitle = {Geostationary {{Satellites Collocation}}},
  author = {Li, Hengnian},
  editor = {Li, Hengnian},
  year = {2014},
  pages = {283--334},
  publisher = {Springer},
  address = {Berlin, Heidelberg},
  doi = {10.1007/978-3-642-40799-4_8},
  urldate = {2025-09-19},
  isbn = {978-3-642-40799-4},
  langid = {english}
}

@article{liu2023Datadriven,
  title = {A {{Data-driven Method}} for {{Realistic Covariance Prediction}} of {{Space Object}} with {{Sparse Tracking Data}}},
  author = {Liu, Hong-Kang and Li, Bin and Zhang, Yan and Sang, Ji-Zhang},
  year = {2023},
  month = jul,
  journal = {Research in Astronomy and Astrophysics},
  volume = {23},
  number = {8},
  pages = {085014},
  publisher = {{National Astromonical Observatories, CAS and IOP Publishing}},
  issn = {1674-4527},
  doi = {10.1088/1674-4527/acd7be},
  urldate = {2024-08-01},
  langid = {english}
}

@article{liu2024Space,
  title = {Space Debris Environment Engineering Model 2019: {{Algorithms}} Improvement and Comparison with {{ORDEM}} 3.1 and {{MASTER-8}}},
  shorttitle = {Space Debris Environment Engineering Model 2019},
  author = {Liu, Yu-Yan and Chi, Run-Qiang and Pang, Bao-Jun and Hu, Di-Qi and Cao, Wu-Xiong and Wang, Dong-Fang},
  year = {2024},
  month = may,
  journal = {Chinese Journal of Aeronautics},
  volume = {37},
  number = {5},
  pages = {392--409},
  issn = {1000-9361},
  doi = {10.1016/j.cja.2023.12.004},
  urldate = {2024-09-28},
  langid = {american}
}

@article{luo2019FocusGEO,
  title = {{{FocusGEO}} Observations of Space Debris at {{Geosynchronous Earth Orbit}}},
  author = {Luo, Hao and Mao, Yin-Dun and Yu, Yong and Tang, Zheng-Hong},
  year = {2019},
  month = jul,
  journal = {Advances in Space Research},
  volume = {64},
  number = {2},
  pages = {465--474},
  issn = {0273-1177},
  doi = {10.1016/j.asr.2019.04.006},
  urldate = {2025-09-21}
}

@article{mains2022IMPACT,
  title = {The {{IMPACT}} Satellite Fragmentation Model},
  author = {Mains, Deanna L. and Sorge, Marlon E.},
  year = {2022},
  month = jun,
  journal = {Acta Astronautica},
  volume = {195},
  pages = {547--555},
  issn = {0094-5765},
  doi = {10.1016/j.actaastro.2022.03.030},
  urldate = {2023-10-15},
  langid = {english}
}

@inproceedings{matney2019NASA,
  title = {The {{NASA Orbital Debris Engineering Model}} 3.1: {{Development}}, {{Verification}}, and {{Validation}}},
  shorttitle = {The {{NASA Orbital Debris Engineering Model}} 3.1},
  booktitle = {International {{Orbital Debris Conference}} ({{IOC}})},
  author = {Matney, M. and Manis, A. and {Anz-Meador}, P. and Gates, D. and Seago, J. and Vavrin, A. and Xu, Y.-L.},
  year = {2019},
  month = dec,
  address = {Sugar Land, TX},
  urldate = {2024-10-08},
  langid = {american}
}

@phdthesis{mei2022Hybrid,
  title = {Hybrid Removal of End-of-Life Geostationary Satellites Using Solar Radiation Pressure and Impulsive Thrusts},
  author = {Mei, Hao},
  year = {2022},
  pages = {173},
  isbn = {9798357551665},
  langid = {english},
  school = {University of Toronto}
}

@phdthesis{nafi2020Practical,
  title = {Practical Optical Survey Strategies for near Geostationary Orbital Debris},
  author = {Nafi, Akhter M.},
  year = {2020},
  school = {Utah State University},
  pages = {199},
  isbn = {9798672180861},
  langid = {english}
}

@article{nasa2025Two,
  title = {Two {{New On-orbit Fragmentations}}},
  author = {NASA},
  year = {2025},
  journal = {Orbital Debris Quarterly News},
  volume = {29},
  number = {1},
  pages = {1},
  langid = {american}
}

@misc{intelsat2024Loss,
  title = {Intelsat Reports {{IS}}-33e Satellite Loss},
  author = {{Intelsat}},
  year = {2024},
  month = oct,
  url = {https://www.intelsat.com/newsroom/intelsat-reports-is-33e-satellite-loss/},
  urldate = {2026-03-18},
  note = {Official statement on the October 19, 2024 anomaly, total loss of the satellite, and failure review board},
  langid = {english}
}

@misc{intelsat2025AnnualReport,
  title = {2024 Annual Report},
  author = {{Intelsat}},
  year = {2025},
  month = mar,
  url = {https://www.intelsat.com/wp-content/uploads/2025/03/2024AnnualReport.pdf},
  urldate = {2026-03-18},
  note = {Official annual report describing service restoration after the Intelsat 33e loss and the related impairment charge},
  langid = {english}
}

@article{rainbow2024Intelsat33e,
  title = {Intelsat 33e Demise Exposes Vulnerabilities in the Space Domain},
  author = {Rainbow, Jason},
  year = {2024},
  month = dec,
  day = {10},
  journal = {SpaceNews},
  url = {https://spacenews.com/intelsat-33e-demise-exposes-vulnerabilities-in-the-space-domain/},
  urldate = {2026-03-18},
  note = {Reports external tracking estimates exceeding 700 debris pieces and notes that Boeing was still investigating the cause},
  langid = {english}
}

@misc{romantisIS33e,
  title = {Intelsat {{IS-33e}} (60{{\textdegree}}E)},
  author = {{Romantis}},
  year = {2026},
  url = {https://www.romantis.com/is33e/},
  urldate = {2026-03-18},
  note = {Satellite information page including manufacturer, platform, mass, and spacecraft image},
  langid = {english}
}

@article{olivieri2023Simulations,
  title = {Simulations of Satellites Mock-up Fragmentation},
  author = {Olivieri, Lorenzo and Giacomuzzo, Cinzia and Francesconi, Alessandro},
  year = {2023},
  month = may,
  journal = {Acta Astronautica},
  volume = {206},
  pages = {233--242},
  issn = {0094-5765},
  doi = {10.1016/j.actaastro.2023.02.036},
  urldate = {2025-09-05}
}

@article{oltrogge2018Comprehensive,
  title = {A Comprehensive Assessment of Collision Likelihood in {{Geosynchronous Earth Orbit}}},
  author = {Oltrogge, D. L. and Alfano, S. and Law, C. and Cacioni, A. and Kelso, T. S.},
  year = {2018},
  month = jun,
  journal = {Acta Astronautica},
  volume = {147},
  pages = {316--345},
  issn = {0094-5765},
  doi = {10.1016/j.actaastro.2018.03.017},
  urldate = {2024-09-07},
  langid = {american}
}

@inproceedings{rossi2009New,
  title = {The New Space Debris Mitigation ({{SDM}} 4.0) Long Term Evolution Code},
  booktitle = {Proceedings of the {{Fifth European Conference}} on {{Space Debris}}, {{ESA SP-672}}, {{CD-ROM}}, {{ESA Communication Production Office}}, {{Noordwijk}}, {{The Netherlands}}},
  author = {Rossi, A. and Anselmo, L. and Pardini, C. and Jehn, R. and Valsecchi, G. B.},
  year = {2009}
}

@article{rossi2016Analysis,
  title = {Analysis of the Consequences of Fragmentations in Low and Geostationary Orbits},
  author = {Rossi, A. and Lewis, H. and White, A. and Anselmo, L. and Pardini, C. and Krag, H. and Bastida Virgili, B.},
  year = {2016},
  month = apr,
  journal = {Advances in Space Research},
  series = {Advances in {{Asteroid}} and {{Space Debris Science}} and {{Technology}} - {{Part}} 2},
  volume = {57},
  number = {8},
  pages = {1652--1663},
  issn = {0273-1177},
  doi = {10.1016/j.asr.2015.05.035},
  urldate = {2022-04-28},
  langid = {english}
}

@article{shu2023Impact,
  title = {Impact {{Risk}} of a {{Debris Cloud}} to {{Spacecraft}}},
  author = {Shu, Peng and Yang, Zhen and Luo, Ya-Zhong},
  year = {2023},
  month = may,
  journal = {Journal of Guidance, Control, and Dynamics},
  volume = {46},
  number = {5},
  pages = {989--997},
  issn = {0731-5090},
  doi = {10.2514/1.G007056},
  urldate = {2023-04-17},
  copyright = {All rights reserved},
  langid = {american},
  lccn = {WOS:000920233000001}
}

@article{shu2025Shortterm,
  title = {Short-Term Evolution and Risks of Debris Cloud Stemming from Collisions in Geostationary Orbit},
  author = {Shu, Peng and Zhao, Meng and Li, Zhen-Yu and Sun, Wei and Li, Yu-Qiang and Luo, Ya-Zhong},
  year = {2025},
  month = mar,
  journal = {Acta Astronautica},
  volume = {228},
  pages = {486--493},
  issn = {00945765},
  doi = {10.1016/j.actaastro.2024.12.016},
  urldate = {2024-12-21},
  langid = {english},
  lccn = {WOS:001393315600001}
}

@article{sun2015Algorithms,
  title = {Algorithms and Applications for Detecting Faint Space Debris in {{GEO}}},
  author = {Sun, Rong-yu and Zhan, Jin-wei and Zhao, Chang-yin and Zhang, Xiao-xiang},
  year = {2015},
  month = may,
  journal = {Acta Astronautica},
  series = {Dynamics and {{Control}} of {{Space Systems}}},
  volume = {110},
  pages = {9--17},
  issn = {0094-5765},
  doi = {10.1016/j.actaastro.2015.01.001},
  urldate = {2025-09-21}
}

@article{wen2024Modeling,
  title = {Modeling {{Medium-Term Debris Cloud}} of {{Satellite Breakup}} via {{Probabilistic Method}}},
  author = {Wen, Chang-xuan and Jin, Zihan and Peng, Chao and Qiao, Dong},
  year = {2024},
  month = aug,
  journal = {Journal of Guidance, Control, and Dynamics},
  volume = {47},
  number = {8},
  pages = {1602--1619},
  publisher = {{American Institute of Aeronautics and Astronautics}},
  issn = {0731-5090},
  doi = {10.2514/1.G008062},
  urldate = {2024-07-29}
}

@article{yao2024Generation,
  title = {Generation of Initial Debris Cloud Distributions for Breakup Events Based on {{CARDC-SBM}}},
  author = {Yao, Tian-Zi and Yang, Zhen and Luo, Ya-Zhong and Lan, Sheng-Wei and Ren, Lei-Sheng},
  year = {2024},
  month = jun,
  journal = {Acta Astronautica},
  volume = {219},
  pages = {580--591},
  issn = {0094-5765},
  doi = {10.1016/j.actaastro.2024.03.060},
  urldate = {2024-07-17}
}

@article{yasaka1992Breakup,
  title = {Breakup in Geostationary Orbit: {{A}} Possible Creation of a Debris Ring},
  shorttitle = {Breakup in Geostationary Orbit},
  author = {Yasaka, Tetsuo and Ishii, Nobuaki},
  year = {1992},
  month = jul,
  journal = {Acta Astronautica},
  volume = {26},
  number = {7},
  pages = {523--530},
  issn = {0094-5765},
  doi = {10.1016/0094-5765(92)90123-Z},
  urldate = {2025-09-21}
}

@article{yasaka1994Remarks,
  title = {Remarks on Orbital Environment Protection at Geostationary Altitude: {{Results}} from Long Term Breakup Simulation},
  shorttitle = {Remarks on Orbital Environment Protection at Geostationary Altitude},
  author = {Yasaka, Tetsuo},
  year = {1994},
  month = oct,
  journal = {Acta Astronautica},
  volume = {34},
  pages = {33--41},
  issn = {0094-5765},
  doi = {10.1016/0094-5765(94)90240-2},
  urldate = {2025-09-21}
}

\end{document}